# 56 Gb/s DMT Transmission with VCSELs in 1.5 um Wavelength Range over up to 12 km for DWDM Intra-Data Center Connects

Annika Dochhan[(1)], Nicklas Eiselt[(1,2)], Robert Hohenleitner[(3)], Helmut Griesser[(4)], Michael Eiselt[(1)], Markus Ortsiefer[(3)], Christian Neumeyr[(3)], Juan José Vegas Olmos[(2)], Idelfonso Tafur Monroy[(2)], Jörg-Peter Elbers[(4)]

[(1)] ADVA Optical Networking SE, Maerzenquelle 1-3, 98617 Meiningen, Germany, dochhan@advaoptical.com
[(2)] Technical University of Denmark (DTU), Dep. of Photonics Engineering, Ørsteds Plads, Build. 343, DK-2800
[(3)] Vertilas GmbH, Daimlerstr. 11d, D-85748 Garching, Germany
[(4)] ADVA Optical Networking SE, Fraunhoferstr. 9a, 82152 Martinsried, Germany

**Abstract** *We demonstrate up to 12 km, 56 Gb/s DMT transmission using high-speed VCSELs in the 1.5 um wavelength range for future 400Gb/s intra-data center connects, enabled by vestigial sideband filtering of the transmit signal.*

## Introduction

Recent development in the field of data center traffic and data center connectivity led to an increased demand of low cost and low complexity high-speed data connections. According to Cisco during the next years, data-center traffic will grow by 25% per year[1]. Recent standardization efforts for 400G intra-data center connections specify link lengths of up to 10 km[2]. Although the standards envisage transmission in the 1.3 µm window with CWDM, for the huge amount of data, DWDM transmission in the 1.5 µm range might be preferable in the future. Recently, we suggested a flexible inter-data center system for 400 Gb/s on four to eight 50-GHz-grid channels[3], depending on the desired reach. In this work, we present an equally scalable technology solution based on a low cost vertical cavity surface emitting laser (VCSEL) as directly modulated transmit laser for reaches of up to 12 km, targeting at intra-data center connectivity. Recent experiments using VCSELs in this wavelength range already showed a reach of several hundred km with 3- and 4-level pulse amplitude modulation (PAM3, PAM4) at 100 Gb/s, but at the expense of introducing coherent detection and polarization multiplexing[4,5]. This effort is prohibitive for low-cost data center environments. Direct detection experiments demonstrated up to 2 km at 56 Gb/s using PAM4[6], and even 100 Gb/s (net rate of ~85 Gb/s) have been transmitted over 4 km using discrete multi-tone transmission (DMT)[7]. By introducing vestigial sideband filtering using the optical multiplexer, which would be present in a DWDM system anyhow, we manage to transmit up to 56 Gb/s DMT over 12 km, which is to the best of our knowledge the longest reach achieved with a VCSEL and direct detection at this rate so far.

## System Setup

Fig. 1 shows the system setup. The DMT signal was generated offline and loaded onto the memory of an 80 GS/s digital-to-analog converter (DAC). A DC bias current was added simultaneously to both parts of the differential signal before it was passed via two electrical probes to the VCSEL die, which was mounted to a ceramic substrate with gold-epoxy. A lensed fiber was used to couple the light into a standard single mode fiber (SSMF). The single mode VCSEL deployed is based on an InP Buried Tunnel Junction (BTJ) design optimized for very high data rates. It features a unique design optimized for low intrinsic parasitics, a short cavity and low threshold current. A cross section and the chip layout of the short cavity VCSEL design is shown in Fig. 2. The device offers a bandwidth of up to 18 GHz, high side mode suppression ratio (SMSR > 45dB), low drive current and power consumption (<25mW). Previous publications with these devices showed up to 56 Gb/s NRZ transmission[8]. To fine-adjust the wavelength, the bias current was varied between 10 and 12 mA, which is a range of optimum bandwidth and extinction ratio. Two different VCSEL samples of the same charge were investigated, with wavelengths of 1522 nm and 1524.5 nm, respectively. These wavelengths are slightly out of the C-band, but the same devices are available for longer wavelengths.

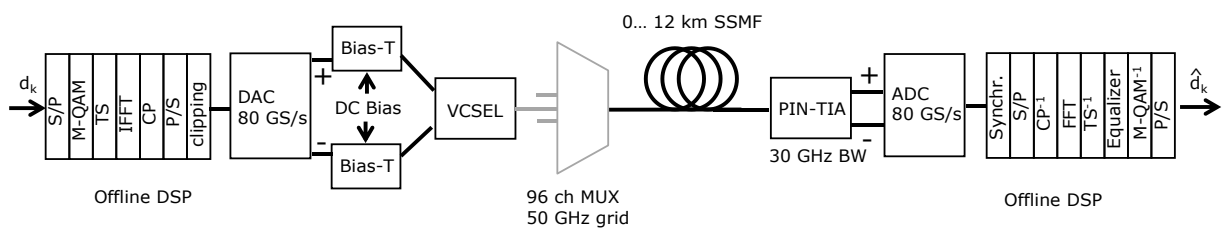

**Fig. 1:** Experimental setup and signal processing steps for the DMT system with VCSEL and optional DWDM multiplexer filter.

The output power was around 1 dBm. The VCSEL was optionally followed by a 96-channel 50-GHz grid DWDM multiplexer and the transmission link of up to 12 km of SSMF. The signal was detected by a 30-GHz bandwidth PIN photodiode integrated together with a transimpedance amplifier (TIA), captured by an 80-GS/s analog-to-digital converter (ADC) and demodulated offline.

For the DMT signal generation, the digital data are mapped into quadrature amplitude modulation (M-QAM) symbols after parallelization (S/P). In a first step, for channel signal-to-noise ratio (SNR) estimation, all subcarriers carry 16QAM. In a second step, bit and power loading is performed using Chow's algorithm[9]. Five training symbols (TS) are added to the 123 data symbols for synchronization and channel estimation. A 512-point IFFT is applied and a cyclic prefix (CP) of 1/64 of the symbol length is added. After serializing, the signal is clipped with an optimized ratio between 9 and 12 dB. At the receiver side, the captured signal is synchronized and converted into parallel data streams. Afterwards, the CP is removed, the FFT is performed and the TS are removed. A decision-directed one-tap equalizer recovers the transmitted data, and after demodulation the bit errors can be counted. The data rate was varied – if an eight channel system is targeted, 56 Gb/s per channel are required, for seven channels, 64 Gb/s are necessary, six channels require 74.7 Gb/s and five channels need 89.6 Gb/s.

**Results and Discussion**

We evaluate the performance with and without the multiplexer. The optimum input power to the receiver was -6 dBm, so without the multiplexer, there is a power margin at the receiver of 4-5 dB. The results are displayed in Fig. 4 (a). Since two samples at different wavelengths are investigated, results are shown for both. The wavelength of VCSEL 1 is at ~1522 nm (results with filled markers), while VCSEL 2 emits at a wavelength of ~1524.5 nm (results with empty markers). Both devices perform quite similar. The maximum reach for a rate of 56 Gb/s is 7.5 km and 64 Gb/s can achieve 5 km. For 74.7 Gb/s the reach is below 5 km, and 89.6 Gb/s cannot achieve the hard-decision FEC limit of 3.8e-3 at all. Next, the multiplexer with a bandwidth of 0.3 nm and an insertion loss of 4 dB was inserted. Previous experiments have shown that vestigial sideband filtering can reduce the effect of chromatic dispersion on the DMT signal significantly[3]. Therefore, we fine-tune the laser bias to shift the laser frequency to an optimum point with respect to the filter passband. In

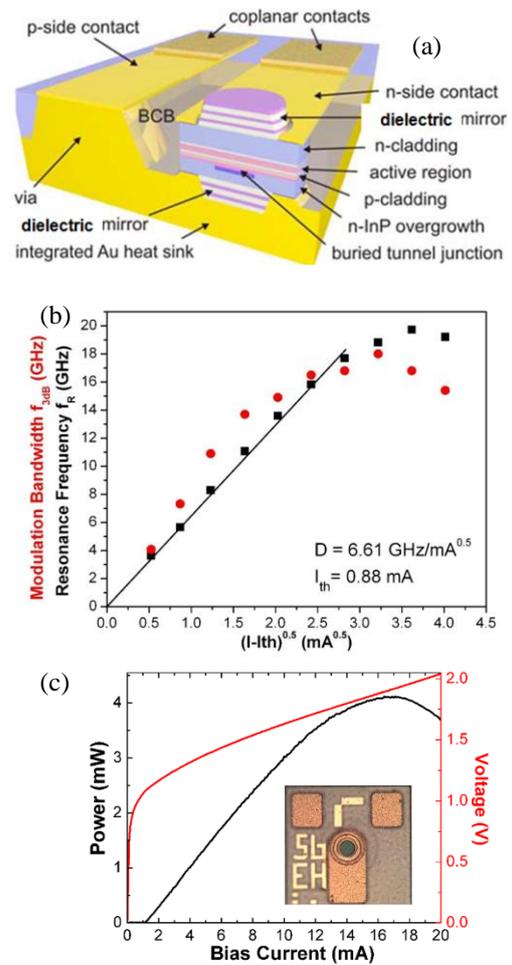

**Fig. 2:** (a) Cross section InP short cavity VCSEL with two dielectric mirrors, (b) Modulation bandwidth and resonance frequency of short cavity VCSEL, (c) LIV and UI characteristics, inset: chip layout.

contrast to our previous experiments, we here have an interplay of the fiber chromatic dispersion, the chirp of the VCSEL and some chromatic dispersion of the filter. The negative influence of chirp on the DMT signal has been investigated in detail by Nishihara[10]. Fig. 5 shows the signal spectrum, captured with an optical spectrum analyzer, as well as the measured filter attenuation and chromatic dispersion within the passband. The signal carrier and the passband center are marked to show the optimum shift of 0.07 nm (~9 GHz) for all considered reaches. There is a large CD variation within the passband, so that part of the DMT signal is effected by a negative dispersion of ~-60 ps and other parts by only ~-10 ps. In fact, this effect might help to enable this long transmission distance in addition to the asymmetrical filtering. Transmission results with filter show a reach of 12 km for 56 Gb/s, 64 Gb/s achieves 10 km and 74.7 Gb/s can be transmitted over 5 km. That means, for a 400G direct detect data center connect superchannel, to cover 12 km, eight

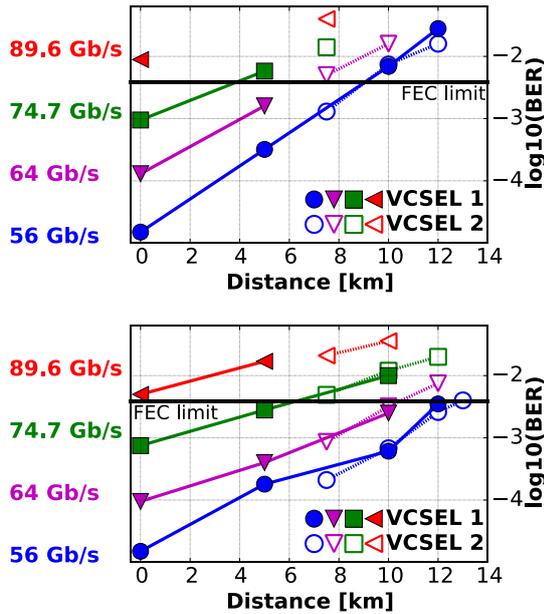

**Fig. 4:** (a) BER vs. transmission reach for various data rates and two VCSEL samples. (a) Without multiplexer filter, (b) With multiplexer filter for asymmetrical (vestigial sideband) filtering.

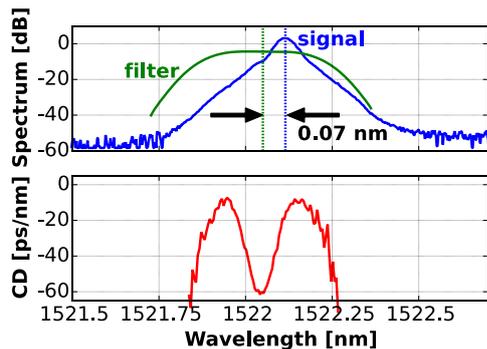

**Fig. 5:** Normalized spectrum of the modulated signal (blue) and filter passband (green) with 0.07 nm optimum shift between passband center and signal carrier. Measured CD within filter passband.

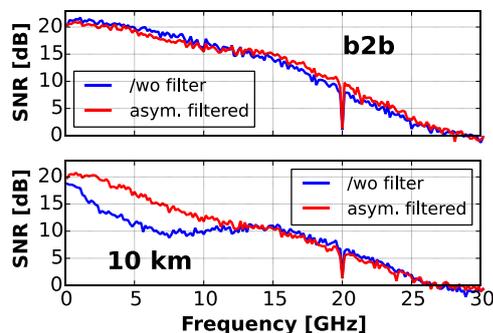

**Fig. 6:** Estimated SNR b2b (upper), after 10 km (lower) with and without multiplexer filter.

DWDM channels in a 50-GHz channel grid are required, while for 10 km this can be reduced to seven channels and for 5 km to even six channels.

The influence of chromatic dispersion and filtering can clearly be observed by evaluating the estimated SNR which is determined in a first channel estimation step to enable optimum bit and power loading. Fig. 6 shows the estimated SNR for b2b and 10 km, with and without the multiplexer filter. The channel capacity is related to the area under the curve, so it can clearly be seen that asymmetrical filtering improves the system. The chirp of the VCSEL increases decay of the SNR, even in the b2b case, although the VSCEL bandwidth is higher than 15 GHz like in Nishihara's results[10].

**Conclusion**

DMT transmission at 56 Gb/s in the 1.5 µm wavelength range is presented over 12 km of SSMF without optical dispersion compensation. This distance is enabled by asymmetrical (vestigial sideband) filtering using a DWDM multiplexer filter. We propose a DWDM system for intra-data center connections with six to eight DWDM channels on a 50-GHz channel grid depending on the desired reach. In a DWDM system, a multiplexer will be present anyhow, so asymmetrical filtering can be performed without the need for additional components.

**Acknowledgements**

The results were obtained in the framework of the SASER-ADVAntage-NET and SpeeD projects, partly funded by the German ministry of education and research (BMBF) under contracts 16BP12400 and 13N13744, and by the European Commission in the Marie Curie project ABACUS.

**References**

[1] Cisco Global Cloud Index. 2014-2019, white paper.
[2] IEEE802.3bs 400 Gb/s Ethernet task force.
[3] A. Dochhan, "Flexible bandwidth 448 Gb/s DMT transmission for next generation data center interconnects", ECOC 2014, Paper P.4.10.
[4] C..Xie, "Generation and Transmission of 100-Gb/s PDM 4-PAM Using Directly Modulated VCSELs and Coherent Detection ", OFC 2014, Paper Th3K2.
[5] C. Xie, "100-Gb/s Directly Modulated VCSELs for Metro Networks", ACP 2014, Paper AF4D.1.
[6] F. Karinou, Directly PAM-4 Modulated 1530-nm VCSEL Enabling 56 Gb/s/λ Data-Center Interconnects", IEEE PTL, Vo.27, No.17, 2015, pp. 1872-1875.
[7] C.Xie, "Single-VCSEL 100-Gb/s short-reach system using discrete multi-tone modulation and direct detection", OFC 2015, Paper Tu2H.2.
[8] D.M. Kutcha, "Error-free 56 Gb/s NRZ Modulation of a 1530 nm VCSEL Link", ECOC 2015, Paper PDP1.3
[9] J. M. Cioffi, "Data Transmission Theory," course text for EE379C (http://www.stanford.edu/group/cioffi/).
[10] M. Nishihara, "Impact of modulator chirp in 100 Gbps class optical discrete multi-tone transmission system", Proc. of SPIE Vol. 8646 86460N-3, 2012.